\documentclass[submission,copyright,creativecommons]{eptcs}
\providecommand{\event}{AiSoS 2013} % Name of the event you are submitting to
\usepackage{breakurl}             % Not needed if you use pdflatex only.
\usepackage{pdfsync}
\usepackage[pdftex]{graphicx}

\title{Systems of Systems Modeled by a Hierarchical Part-Whole State-Based Formalism}
\author{Luca Pazzi
\institute{University of Modena and Reggio Emilia\\
DIEF-UNIMORE\\
Via Vignolese 905, I-41125 Modena, Italy}
\email{luca.pazzi@unimore.it}
}
\def\titlerunning{Systems of Systems Modeled by a Hierarchical Part-Whole State-Based Formalism}
\def\authorrunning{Luca Pazzi}
\begin{document}
\maketitle

\begin{abstract}
The paper presents an explicit state-based modeling approach aimed at modeling Systems of Systems behavior.  The approach allows to specify and verify incrementally safety and liveness rules without using model checking techniques.  The state-based approach allows moreover to use the system behavior directly as an interface, greatly improving the effectiveness of the recursive composition needed when assembling Systems of Systems.
\end{abstract}

\section{Introduction}
%intelligent buildings, smart cities, transport systems, etc.
%There is a need for new modeling formalisms, analysis methods and tools to help 
%make trade-off decisions during design and evolution avoiding leading to sub-optimal 
%design and rework during integration and in service. The workshop should focus on 
%the modeling and analysis of System of Systems.

% The need of controlling the overall complexity of modern infrastructures calls for new conceptual methods, although the concept of system is present in everyday life, 

While traditional systems engineering focuses on systems made of simple constituent parts, Systems of Systems (SoS) comprise multiple autonomous systems which can be very different in technology, context, operation, geography and conceptual frame~\cite{SoS1}. The coordinated behavior of such systems constitutes the primary behavior of the Sos itself. Finally, Sos have a recursive nature, each component of a Sos being possibly a SoS itself.

Although the difference can be at first sight very loose, since constituent parts in traditional system engineering are often system themselves, engineering Systems of Systems poses very specific challenges due to the heterogeneous nature and role of the systems participating in the whole assembly.
In other words, the focus shifts from choosing the right system to choosing the system, or even multiple systems, able to satisfy the right specific behavioral and functional requirements.  Component systems in SoS need therefore to be easily interchangeable both in the design and in the operation phases. 

Such an heterogeneous diversity and interchangeability context calls for a unifying language for describing and prescribing the behavior of both the components and the assembled system.  Such a language should be general enough for the sake of taking into account system diversity, but, at the same time, it should be able to express modal and logical properties of the global system being engineered, that is what the system behavior should or should not be allowed to do.
Such a language should moreover take into account architectural issues. 
%in order to allow loose coupling among systems being assembled in more complex systems.  

The recursive nature of the SoS approach, and the need for interchangeability of component parts while still satisfying requirements, calls in fact for thinking system architecture in a modular way.  A system must be modeled by a module which should be able to play different roles in different compound systems: this in turn requires that a system should not be allowed to know any detail of the compound systems that will contribute to form. On the other hand, the compound system should be able to know in detail the behavior of the single systems by which it is composed by.
% employs in order to make its own global behavior. 
The relationships among component and compound system is therefore asymmetrical, distinguishing and clearly identifying the part from the whole.  The whole is required to know its parts, the parts are forbidden to know the whole in order to be interchangeable among different compound systems.  It is finally remarked that both roles must be played by the same module, since, as observed, each system is a SoS itself.

The asymmetric part-whole composition framework suggests consequently, an asymmetric communication framework.  
The compound whole should be able to control \emph{directly} the component parts: the parts, on their turn, should not control directly the whole, but be allowed to influence it only \emph{indirectly}.  Consider for example an Air Traffic Control, and suppose to model it as a SoS where airplanes are, among the others, the principal system components. While an ATC may issue commands to the different planes being under its control, airplanes may only \emph{notify} ATC of their position, altitude, possible failures as well as requests for landing, approaching, takeoff, and so on.  ATC will consider each notification or request from one of its subsystems (planes, runaways, safety ground systems, weather stations) and issue back commands to them taking into account the \emph{global state} of the compound ATC system, resulting from the different planes position and altitude, weather conditions, runaway free or in use by other planes, safety ground systems, and so on. 

\begin{figure}[h]
   \centering
   \includegraphics[scale=.43]{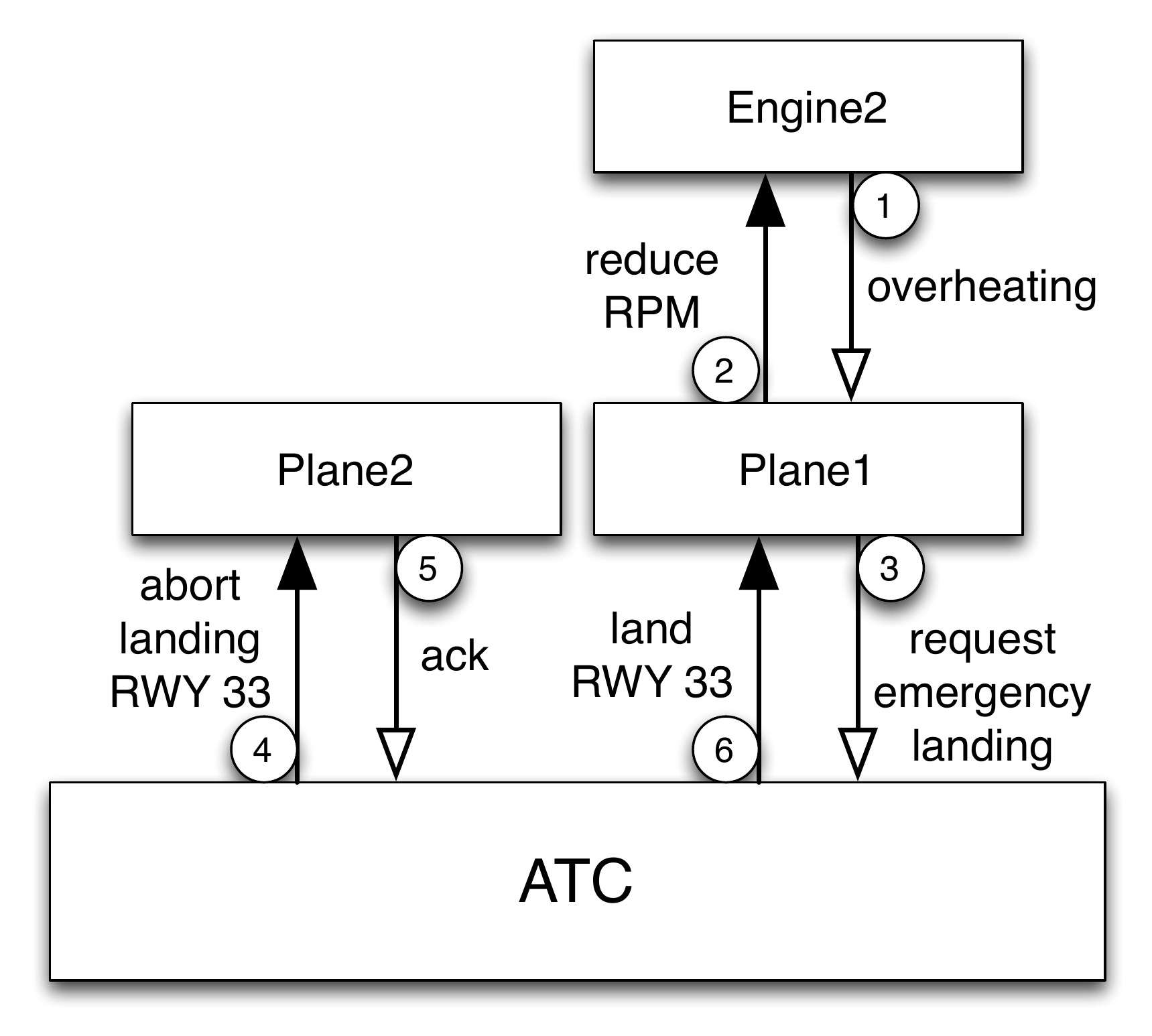}
   \caption{Event flow exchanged amongst four systems belonging to an Air Traffic Control (ATC) scenario.  Command and feedback among systems are represented, respectively, by black and white arrows.}
   \label{atcEventFlow}
\end{figure}

Figure~\ref{atcEventFlow} shows a typical flow of events from different systems arranged hierarchically relating to an ATC scenario.  An engine ($\mathtt{Engine2}$) notifies the control system of the airplane to which it belongs an overheating failure (1). The airplane $(\mathtt{Plane1}$) reacts to the failure by~(2) commanding $\mathtt{Engine2}$ to reduce power and by~(3) notifying the $\mathtt{ATC}$ of the problem.  Observe that the system $\mathtt{Plane1}$ has the engine under its direct control, but it can not send commands directly to the $\mathtt{ATC}$ system, but only notifications.  The $\mathtt{ATC}$ system, on its turn, has $\mathtt{Plane1}$ and $\mathtt{Plane2}$ under its control, and gives a command to the second plane to abort landing~(4) on specified runaway.  Once the $\mathtt{Plane2}$ acknowledges to abort its landing~(5), the runaway is free for $\mathtt{Plane1}$, which is given the command~(6) to undertake its emergency landing.

\subsection{Structure of the work}

By the approach proposed in this paper, a System of Systems (i) has control over other systems and (ii) is in turn controlled by other systems.  
% A single system must therefore take into account the \emph{two} different roles, which have necessarily to agree. 
According to such a view, its behavior has to play seamlessly both roles, that is, it has to be be, \emph{at the same time}, a controller and a controlled behavior.  The system behavior has moreover to agree within the architectural and event flow framework depicted above.

We discuss the 
%three 
different aspects in the rest of the paper.  
In Section~\ref{implicitVsExplicit} we argue that interacting systems can be modeled equivalently by introducing an explicit additional system having the original systems as components, which encapsulates the dynamical aspects regarding  interaction among the original systems.  We argue moreover that such a modeling brings advantages in software quality terms. 
(It is therefore implied that any set of interacting processes can be modeled by a more effective SoS having the original interacting systems as components.)
In Section~\ref{explicitPWS} we we show the feasibility of the approach hypothesized in the previous Section by adopting the PW-Statecharts state-based formalism, which allows to represent, by a single construct, the behavior of a system acting both as whole and as part of more complex wholes. We show moreover that the state semantics of the system acting as whole is computable and that it is possible to check its correctness against safety and liveness rules by exploring a finite state diagram.

%Holons are discussed in Section~\ref{holonsFramework}. (We show that PW-Statecharts state-based formalism allows to deploy a system behavior which acts both as an interface, once used as a part, and as a system control, once used as a whole.)

\section{Implicit approach in system modeling}
\label{implicitVsExplicit}

We propose to use state diagrams for expressing behavioral specifications of Systems of Systems since state-based modeling is clear, realistic, formal and rigorous~\cite{Harel:ICSE:96} 
for describing and prescribing the behavior of both the components and the assembled system. Such a language is general enough for taking into account system diversity and for enforcing and verifying, through model-checking techniques, modal and logical properties of the global system being engineered. In this section we explore the relationship between state-based behavioral descriptions and architectural issues.

Modular encapsulation of state-based behavioral abstractions is still an open issue. Object-oriented development methodologies, such as Real-time~UML~\cite{Douglass:1998:RTU}, encapsulate the state behavior of single systems within state modules hosted into parallel Statecharts~\cite{Harel:87} sections.

A system is thus modeled by a set of interacting parallel state machines (each state machine hosted within an AND-decomposed state, each single state of the machine being and XOR-decomposed state), which synchronize through message exchange and mutual condition testing. Statecharts state decomposition mechanism furnishes thus a straightforward way of representing single entities, which compose into more complex systems through synchronization.  In other words, process synchronization denotes system aggregation.

\begin{figure}[h]
   \centering
   \includegraphics[scale=.5]{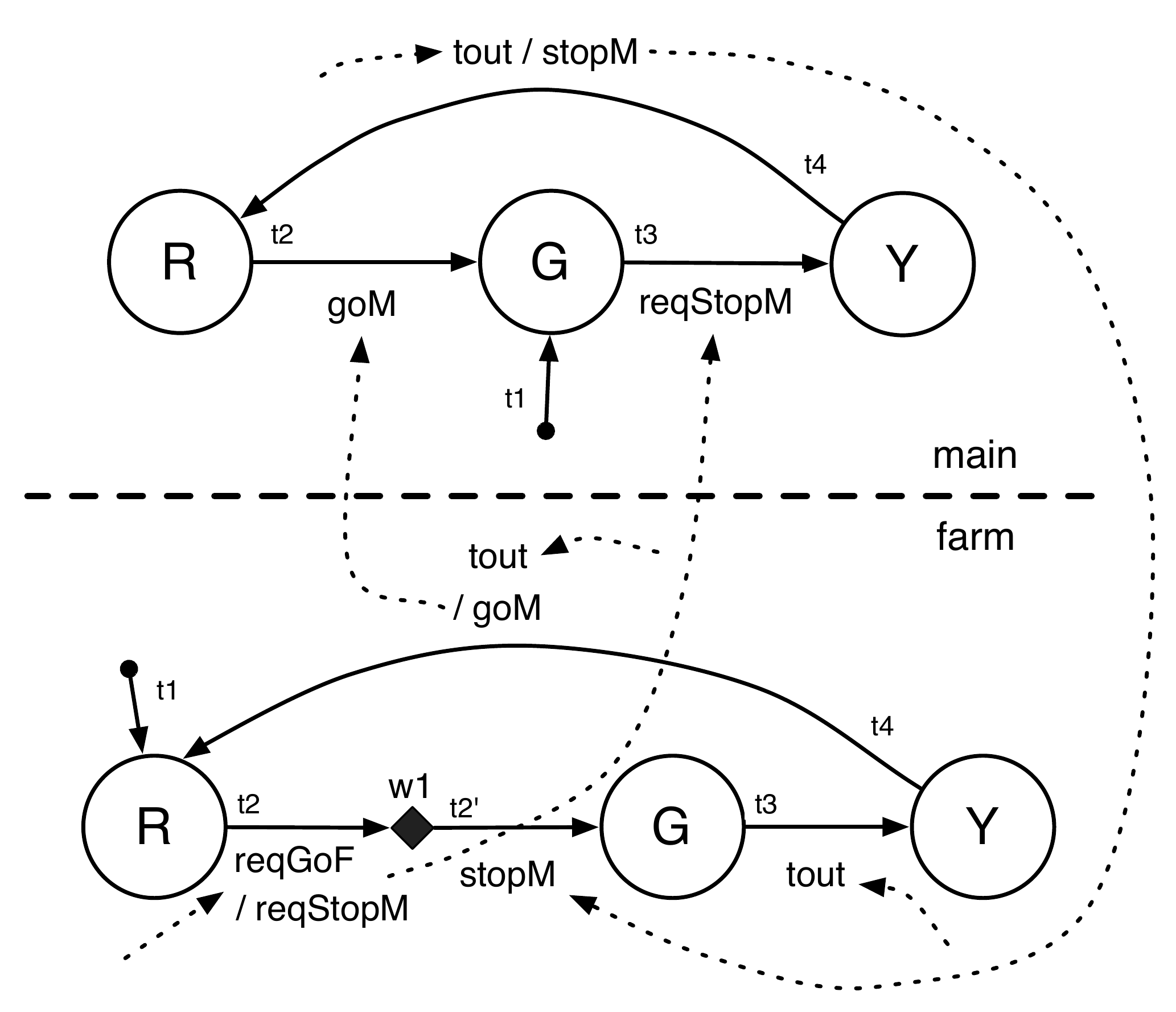}
   \caption{Two mutually interacting parallel state-based processes, each associated to a traffic light regulating the access from a farm to a main road.  Dotted arrows show mutual interactions among the two state machine by direct event forwarding.  
% Squares embedding a $T$ denote timed states.  
Timeout ($\mathtt{tout}$) events come from timer state machines (not shown).
   }
   \label{communicatingTLs}
\end{figure}

\paragraph{Example.} Figure~\ref{communicatingTLs} shows two interacting state machines, $\mathtt{farm}$ and $\mathtt{main}$, each hosted within a Statecharts' parallel sections.  Each can be seen as a state-based process.
%
% The controllable behavior of a system has to be the \emph{same} behavior which controls further systems.
The farm road is normally stopped, while the main road is normally open. States $\mathtt{R}$, $\mathtt{G}$ and $\mathtt{Y}$ stand for lights read, green and yellow. A car arriving at the crossroad from the farm road is sensed, trough some device not in the example, by the $\mathtt{farm}$ traffic light, which asks the $\mathtt{main}$ to block the main road by sending it an ``open'' request ($\mathtt{reqGoF}$), which in turn sends a ``stop'' request to the main-troad traffic light ($\mathtt{reqStopM}$). The farm traffic light moves then to a special wait state $\mathtt{w1}$, aimed at modeling the fact that we have to wait for the main road traffic light to go to the $\mathtt{R}$ state before moving to the $\mathtt{G}$ state.

---

The two state-based processes of the example in Figure~\ref{communicatingTLs}, once synchronized, become a single process, which in turn denotes a single system, namely \emph{the crossroad controller system}.  
It can be observed that process synchronization can be achieved by two different approaches: by direct communication among system components, as in the two traffic lights example of Figure~\ref{communicatingTLs} or through an explicit additional entity representing the system being modeled, which has the system components as parts and hosts the system behavior as a whole. The two approaches have been named respectively \emph{implicit} and \emph{explicit} system modeling~\cite{pazzi:99:DKE}. The explicit approach will be discussed in Section~\ref{explicitPWS}.

Focusing on system components, a practice inspired by real-world observation and experience, may be misleading at the system level.
A physical system is in fact assembled from a set of physical components, which exercise physical control one upon another.
For example, a set of mutually related devices may globally exhibit a systemic behavior through  direct physical interactions, which cause, in turn, state changes in related components. 
% Chains of component state changes are at the basis of the global system behavior. 
However, a different view is possible, since the global state changes resulting from a chain of causally induced state changes at the component level can be seen as a \emph{single} state change at the system level.
Consequently, a number of state transitions at the component level may be represented by a single state transition at the system level. In the same way, the global system behavior which implicitly results from direct interactions may be explicitly represented in the model.

Most programming and modeling paradigms are committed towards the implicit approach in modeling system behavior, since they mimic the physical interactions among components by direct event messages, as observed.  Such a commitment towards the implicit modeling of systems has major drawbacks. For example mutually interacting processes lack clarity and understandability, since they have to embed synchronization details in the code, as shown in the two traffic light example of Figure~\ref{communicatingTLs}.  Resulting modular abstractions are therefore not self-contained and tightly-coupled~\cite{meyer:88}.  Moreover, it may be the case that they have to embed behavioral details which pertain to the overall systemic behavior, as in the traffic light example where the farm traffic light has to introduce a wait state $w1$ which, in addition to its own state $Y$(ellow), order to model the switchover timing.

Consequently, the implicit global behavior is difficult to understand, modify, reuse, extend, and so on. 
State machines, in the Statecharts variant, lack moreover a definite and precise state semantics~\cite{vonderBeek:vB:94}, that is, it is not possible to establish in advance if and when, and in which global state, interacting state machines stop. 
%
%It cannot be determined in advance, in fact, when and if, and in which global state, mutually communicating and synchronizing state machines halt.  
%%
%Behavioral aspects of parallel systems made of behaviorally independent communicating parts can not be  verified against safety and liveness logical statements. 
%%
%The behavior of the globally assembled system can only be verified through model checking techniques, which can be performed however only at a later time.  

\section{Explicit modeling: Part-Whole Statecharts}
\label{explicitPWS}

In order to overcome the problem observed, we propose a model of concurrent autonomous systems where composition is restricted by a part-whole hierarchy: a System of Systems is modeled by a central controller system, referred to in the following of the paper as the ``whole'', which has one or more controlled system as its components, called ``parts''.  The behavior of each system is specified by extended state machines through a special state-based language, Part-Whole Statecharts~\cite{pazziACSD12} which is able to represent, by a unique state diagram, the behavior of the system seen both as whole having other systems as parts, and as a system being part of other wholes.  
Such extended state machines are able to process both the different kinds of events (commands and notifications exchanged between the system acting as whole and the systems acting as components). 
%
%Such an extended state machine has to model both the behavior of the system acting as a whole, and at the same time, of the (same) system acting as a part.  

Although we do not report here the complete syntax of PWSs in this paper (the reader may refer to~\cite{pazziACSD12} for full details), we illustrate the main features of the approach by the following example.

\begin{figure}[h]
   \centering
   \includegraphics[scale=.33]{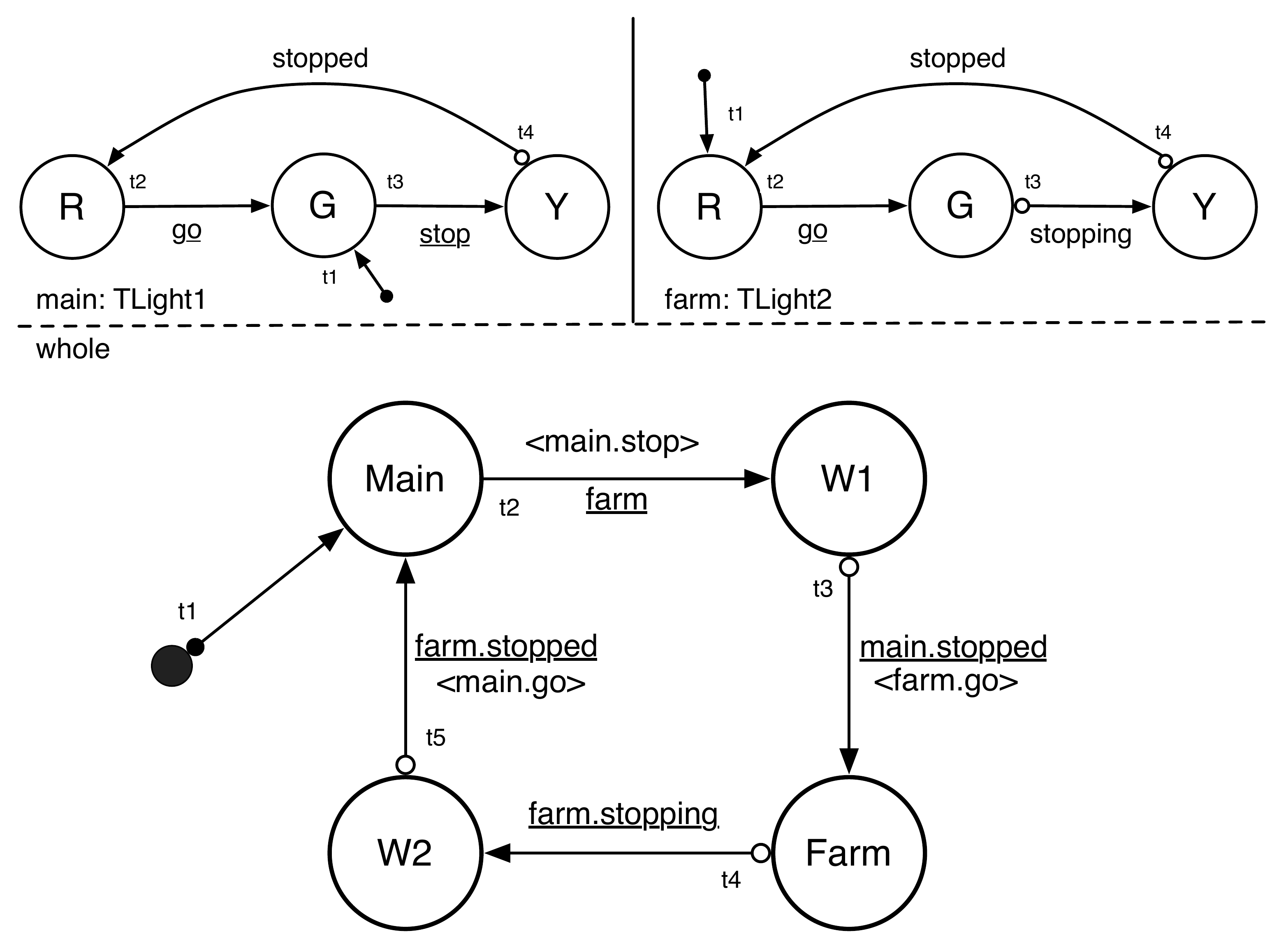}
   \caption{Explicit synchronization by Part-Whole Statecharts (adapted from~\cite{pazziACSD12}). }
   \label{CrossPlusVerification2}
\end{figure}

\paragraph{Example.}  The two synchronized traffic lights of the previous Section may be equivalently modeled by the Part-Whole Statechart of Figure~\ref{CrossPlusVerification2} representing the crossroad system as a whole.  A PWS is basically constituted by two communicating sections, the ``whole'' and the ``assembly'' sections, separated by the dotted horizontal line. The ``whole'' section hosts a state diagram which \emph{explicitly} coordinates a set of state machines hosted in the ``assembly'' section and allows to view the semantics of interactions amongst participating systems at a glance. 
Any interaction among the state machines in the assembly section is forbidden, in order to force the designer to make explicit the synchronization semantics, which is indeed ``shifted'' to the whole section.
State diagrams in the upper section of the PWS are called \emph{state interfaces}.  The lower section diagram embeds synchronization details which bind the whole to the components' interfaces. The whole section of the PWS may use only what is present in the such interfaces, that is states, event and triggers labeling transitions.
State transition $t_{3}$ from $\mathsf{G}$ to $\mathsf{Y}$ of state machine $\mathsf{main}$, for example, may be triggered by event $\mathsf{stop}$: consequently, transition $t_{2}$ in the whole section will trigger such a transition by having the command event $\mathsf{main.stop}$ in its own command list.
The ``whole'' state diagram becomes, in turn, the state interface of the modeled crossroad system which has the two original state machines as parts. 
%(Figure~\ref{CrossPlusVerification2Interface}).
%
State changes and event coming from the assembly trigger state transitions in the whole section of the PWS, which in turn send back action events to the state machines in the assembly.  
For example, transitions $t_{3}$ reacts to the notification event $\mathsf{stopped}$ coming from the components traffic light $\mathsf{main}$ through the trigger $\mathsf{main.stopped}$, which triggers the transition and which sends, in turn, command event $\mathsf{go}$ to component $\mathsf{farm}$ through the action $\mathsf{farm.go}$.  
It can be observed that, while the ``whole'' moves from state 
$\mathsf{W1}$ 
to state 
$\mathsf{Farm}$ 
the assembly of state machines moves, accordingly, from the global state
$(\mathsf{Y},\mathsf{R})$
to the global state
$(\mathsf{R},\mathsf{G})$
thus furnishing a first base towards the computation of state semantics discussed in Section~\ref{stateSemantics}.

%\begin{figure}[h]
%   \centering
%   \includegraphics[scale=.33]{CrossPlusVerification2Interface.pdf}
%   \caption{The interface obtained from the PW-Statechart of Figure~\ref{CrossPlusVerification2}.  Component traffic lights, as well as details pertaining them, are hidden.  The behavior starts in state $\mathsf{Main}$, then transition $t_{2}$ be triggered by event $\mathsf{farm}$, then the system performs a cycle through the other states by the autonomous state transitions $t_{3}$, $t_{4}$ and $t_{5}$.  }
%   \label{CrossPlusVerification2Interface}
%\end{figure}

\subsection{State semantics}
\label{stateSemantics}
It is possible to determine \emph{at design time} the state configurations the set of system components, referred to int he rest of the paper as \emph{assembly} of components, will assume when the control is in a given state of the state machine which controls the behavior of the compound system~\cite{pazziACSD12}.  

By \emph{state configuration} it is meant a tuple 
$\pi=\langle q_{1}, q_{2}, \ldots, q_{N} \rangle$, 
where $N$ is the number of systems in the assembly of components, and $q_{i} \in Q_{i}$, the set of states of the $i$-th system in the assembly, with $i \in N$. 
Let each state configuration $\pi$ denote trivially the \emph{basic} proposition 
``the assembly of systems is in configuration $\pi$''
about the global state of the assembly of systems.  

A \emph{state proposition} $s$ is a disjunction of basic state propositions 
$s = \pi_{1} \vee \pi_{2} \vee \ldots \vee \pi_{k}$
Alternately, a state proposition can be seen as a set of possible configurations of the assembly of system components, i.e.~
$s = \{ \pi_{1}, \pi_{2}, \ldots, \pi_{k} \}$.

Let $A$ be one of the states of the state machine $W$ which controls the behavior of the compound system. Let $\mathsf{sem}(A)$ denote a state proposition, called the \emph{state semantics} of $A$.  
The state semantics of each state $S \in Q_{W}$ in the state machine $W$ can be computed inductively by following the state diagram structure.  

\begin{figure}[h]
   \centering
   \includegraphics[scale=.33]{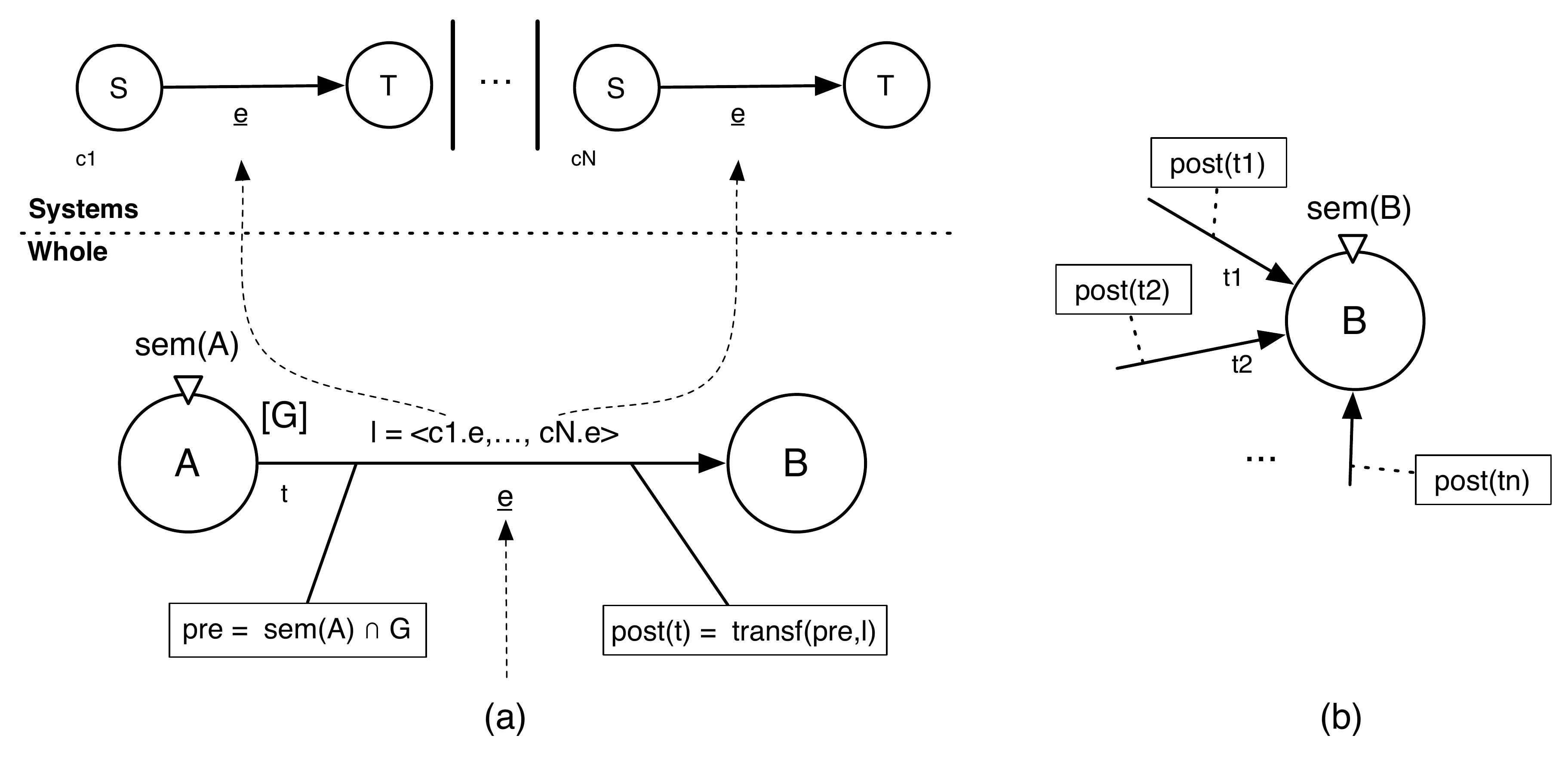}
   \caption{State semantics determination: basic state transition case.  In~(a) it is shown the how the set of allowed assembly configurations are determined for a single transition, in~(b) it is shown how the full state semantics is determined for a state having $N$ incoming state transitions. }
   \label{semCalculusAiSoS2}
\end{figure}

Let us suppose a state transition $t$ links state $A$ and $B$ (as in Figure~\ref{semCalculusAiSoS2}) and that, by the induction hypothesis, the state semantic of the starting state $A$ of the transition is known. 
Let $l = \langle a_{1}, a_{2}, \ldots, a_{k} \rangle$ be a list of action commands directed towards the $N$ systems $c_{1}, c_{2}, \ldots, c_{N}$ making the assembly.  Each command action $a \in l$ is of the form $c_{i}.e$ meaning that system $c_{i}$ has a state transition which can be triggered by event $e$. Finally, state proposition $G$ within square brackets acts like a guard condition in ordinary Statecharts, that is it must hold in order for the transition to be taken.  

Let us suppose the current configuration of the assembly of component systems be $\pi_{c}$ when the current state of the controller is $A$. Then either $\pi_{c} \in G$ or not.  Since, only in the former case, the transition is triggered, and since, by the inductive hypothesis $\mathsf{sem}(A)$ is known and holds of the current global state of the assembly, then $\pi_{c}$ is such that transition $t$ is triggered if and only if it belongs to the set of configurations 
$\mathsf{pre}=\mathsf{sem}(A) \cap G$.

Let $I(l)$ be the set of indexes such that $i \in I(l)$ iff  $c_{i}.e \in l$.
In case transition $t$ is triggered, the action commands $c_{i}.e$ in $l$ prescribe state transitions $q_{i} = \delta(q_{i}, e)$ in component $c_{i}$ of the assembly, with $i \in I(l)$.
Let $\pi_{c}=\langle q_{1}, q_{2}, \ldots, q_{N} \rangle$ be the current state configuration.  Then $\pi_{c}$ is transformed into the tuple $\pi'_{c}$ in such a way that each $q_{i}$ in tuple $\pi_{c}$ is replaced by $q'_{i}$.  
We denote the transformation induced by the command list $l$ on the assembly configuration $\pi_{c}$ by function $\mathsf{transf}$, such that $\pi'_{c}=\mathsf{transf}(\pi_{c},l)$.  We define $\mathsf{transf}$ equivalently for sets of assembly configurations $\Pi$, meaning that $\Pi'=\mathsf{transf}(\Pi,l)$ iff for any $\pi \in \Pi$ we have that $\mathsf{transf}(\pi,l) \in \Pi'$.

Given a set of state configurations which hold when the system controller is in state $A$, the set of configurations which hold for the arrival state of the transition $t$ is then given by:

\begin{equation}
\label{e1}
\mathsf{post}(t)=\mathsf{transf}(\mathsf{sem}(A) \cap G,l)
\end{equation}

which can be meant as the state semantics of state $B$ induced by state transition $t$.
Since state $B$ may have different incoming transitions, its full semantics, that is the entire set of configurations the assembly may assume when the controller is in state $B$, is given by the ``union'' of the different incoming state transition semantics $\mathsf{post}(t)$:

\begin{equation}
\label{e2}
\mathsf{sem}(B)=\bigcup_{t \in i(B)} \mathsf{post}(t)
\end{equation}

where $i(B)$ denotes the set of state transitions which have $B$ as arrival state.  Finally, the base case on which the inductive hypothesis is grounded is that the semantics of the initial state of the whole is given by a configuration which contain the tuple of the initial states of each components in the assembly.
 
\begin{figure}[h]
   \centering
   \includegraphics[scale=.33]{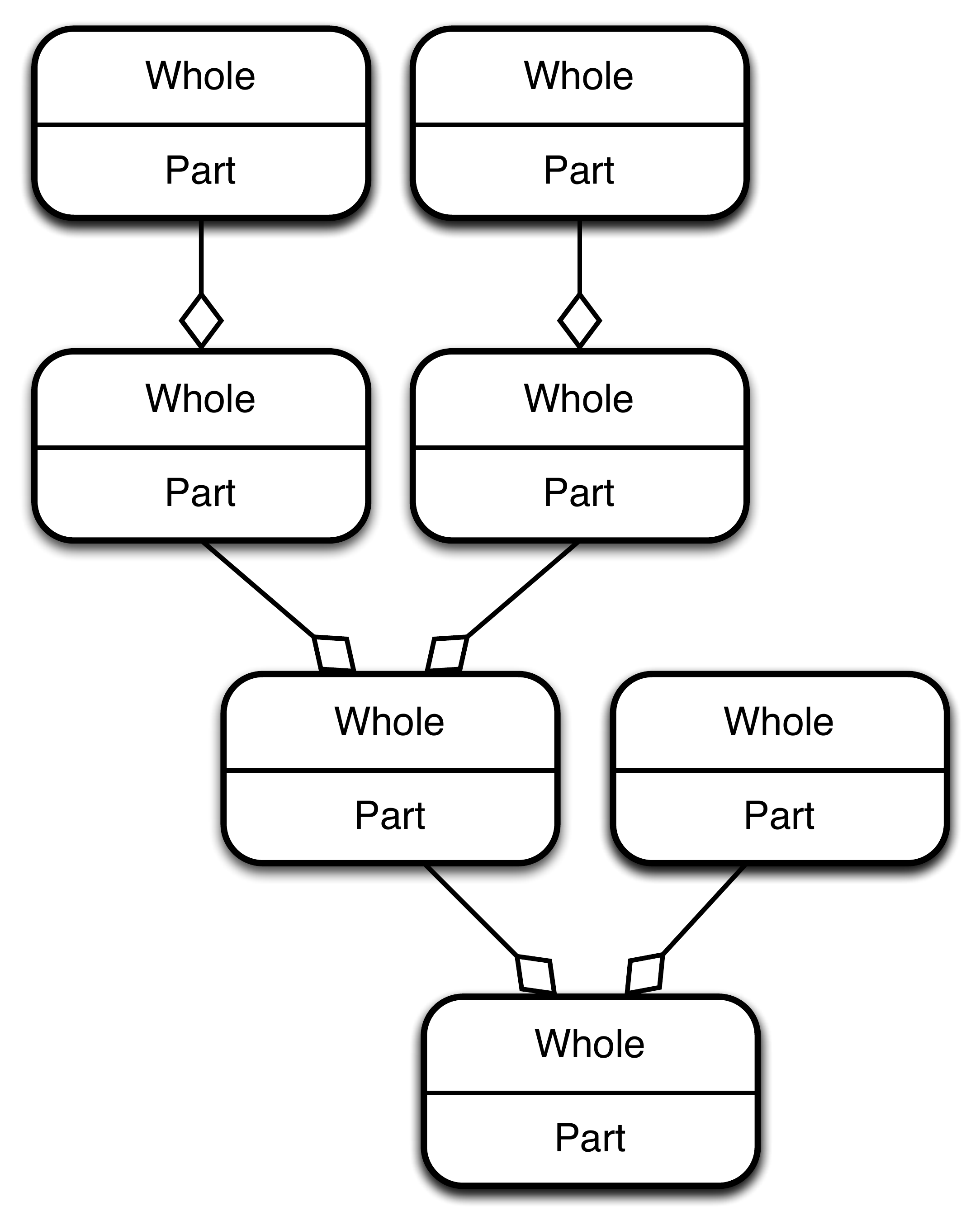}
   \caption{A holarchy, i.e.~a part-whole hierarchy of holons.}
   \label{holons2}
\end{figure}

\subsection{Formal safety verification}

Safety and liveness issues are raised by systems competing for a mutually exclusive resource.  Each behavioral process associated with the system has to check whether the resource is free, and in case it is not, it has to ask the other processes to release it.  On its turn, a process holding a resource should release it eventually.  The two airplanes in Figure~\ref{atcEventFlow} compete for the same resource, that is the runaway. It may be observed that the ATC task is to ensure that no two airplanes have the same runaway in use at the same time; it may be observed also that different airplanes may also communicate directly one with the another, as normally happens in small airfields where no central control is available.

The implicit, traditional approach, uses model checking techniques in order to explore all the feasible mutual behavior in order to check whether rules are satisfied, for example:

\begin{enumerate}

\item the system starts in the global state $(\mathsf{G},\mathsf{R})$ such that only one process is in the critical section (i.e., only one road has access to the crossroad);
\label{firstSafetyRule}

\item it is always guaranteed that it is never the case that both processes are in the critical section, (i.e, is the global state $(\mathsf{G},\mathsf{G})$ is not reachable);
\label{mainSafetyRule}

\item each process is guaranteed to release the critical section (i.e., each traffic lights moves from $\mathsf{G}$ to $\mathsf{Y}$ then $\mathsf{R}$).
\label{thirdSafetyRule}

\end{enumerate}

The main advantage in having an explicit representation of behavior is that the system will always be in a finite set of states \emph{and in no other state}. Each state of the whole section can be put in correspondence with a set of allowed configurations of states of the components' assembly, as shown in Section~\ref{stateSemantics}.  It is then possible to trivially visit the finite state diagram in the whole section in order to check whether safety and liveness rules are satisfied.  In the crossroad example we have for example:

\begin{eqnarray}
 \mathsf{sem(Main)} & = & \{ (\mathsf{G},\mathsf{R}) \}\\
 \mathsf{sem(W1)} & = &   \{ (\mathsf{Y},\mathsf{R}) \}\\
 \mathsf{sem(Farm)} & = & \{ (\mathsf{R},\mathsf{G}) \}\\
 \mathsf{sem(W2)} & = &   \{ (\mathsf{R},\mathsf{Y}) \}
\end{eqnarray}

Hence rules~\ref{firstSafetyRule},~\ref{mainSafetyRule} and~\ref{thirdSafetyRule} can be trivially verified to hold for the crossroad PWS.  Due to the overall compositionality of the approach, the verification process is moreover incremental and fully compositional.  Once the crossroad is \emph{safe}, it can be composed into further systems having it as part, without the need to reconsider its internal safety.

\begin{figure}[h]
   \centering
   \includegraphics[scale=.33]{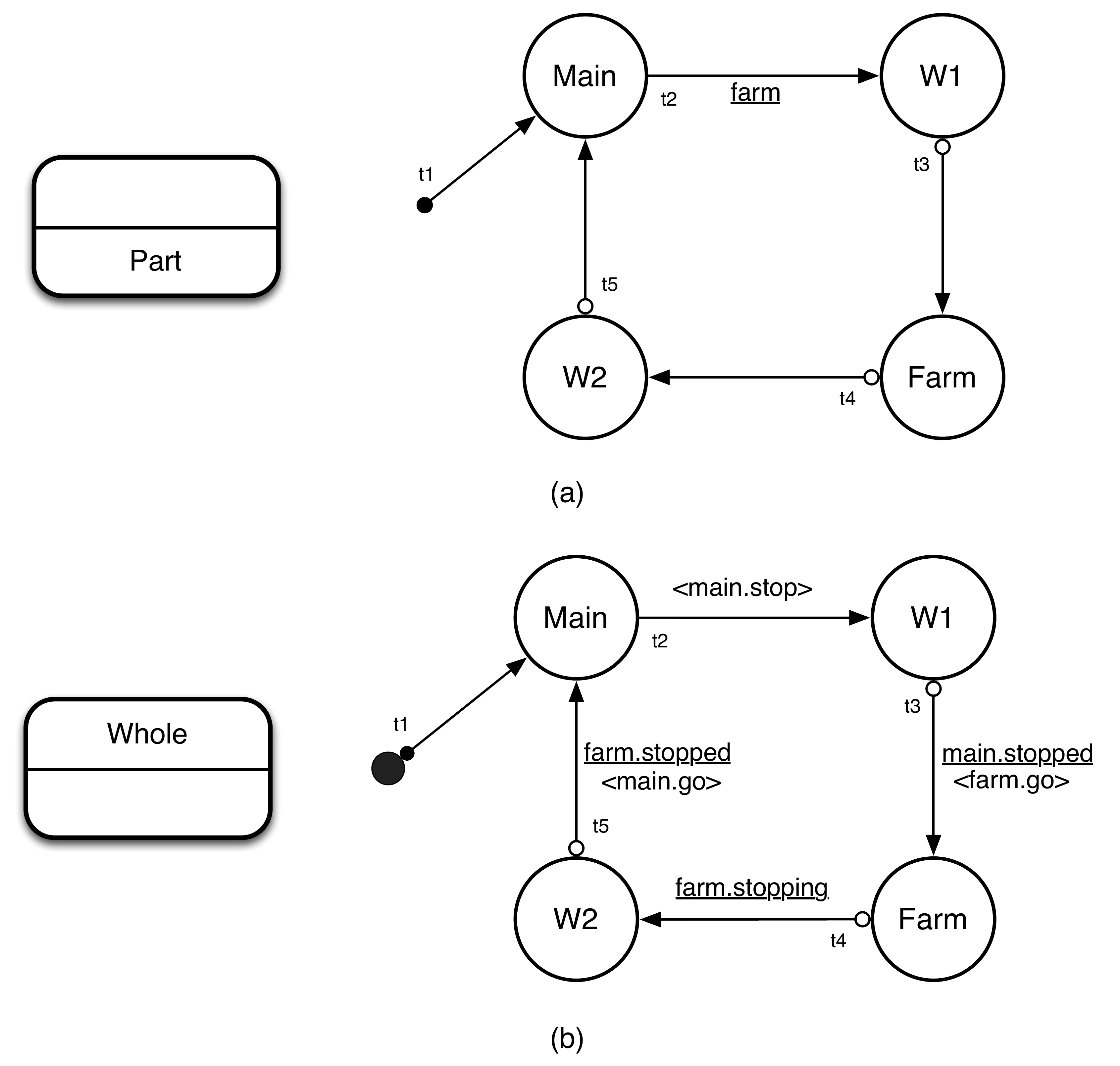}
   \caption{The interface section of an holon~(a) can be obtained from the whole automaton~(b) by stripping implementation details.}
   \label{partVsWhole}
\end{figure}

\begin{figure}[h]
   \centering
   \includegraphics[scale=.23]{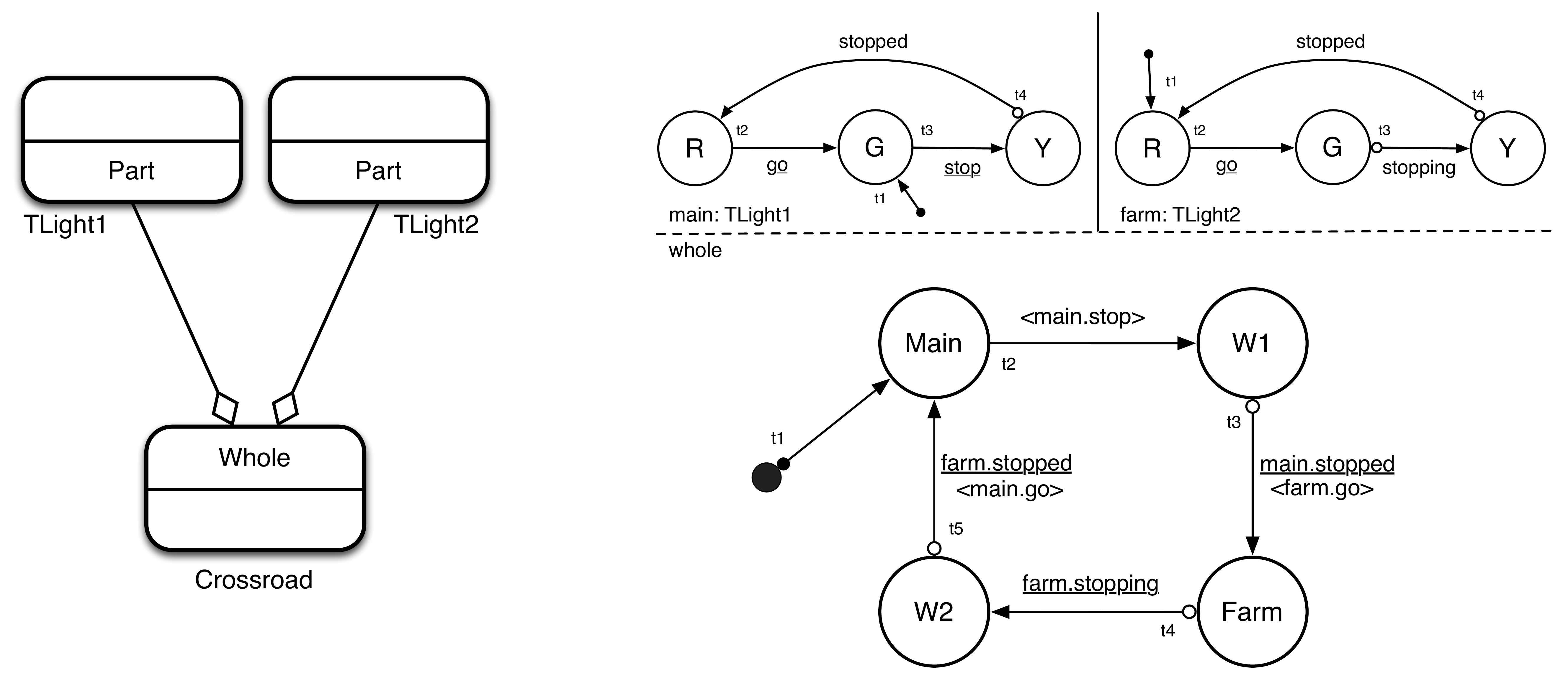}
   \caption{Basic holonic part-whole composition from a single PWS.}
   \label{partVsWhole2}
\end{figure}

%\begin{figure}[h]
%   \centering
%   \includegraphics[scale=.23]{partVsWhole2.pdf}
%   \caption{Basic holonic part-whole composition from a single PWS.}
%   \label{partVsWhole2}
%\end{figure}

\section{Conclusions}

Interacting state machines synchronize system behavior by message exchange. Such messages, however, denote different kinds of information. Typically, systems communicate either by ``peer to peer'' or ``part to whole'' message exchange, the latter case pertaining to systems composed of other systems.  The problem consists, at the ontological level, in determining whether two systems stand in the former or in the latter relationship.  Statecharts, for example, do not distinguish amongst the two cases.
%, which have a different composition semantics. 

As observed, vertical, part-whole, system composition is asymmetrical in nature and preserves model reusability. On the other hand, horizontal, peer to peer message exchange hinders model reusability, since it forces system modelers to introduce exogenous details within systems being modeled, bringing severe limitations to the overall software quality of the modeled systems.

Physical interactions in physical systems denote in fact less evident conceptual structures, which host the overall interaction and synchronization knowledge among the component parts. By introducing additional system entities with the aim of hosting such knowledge in a localized and compact manner, we obtain a part-whole hierarchy of systems, called holarchy~\cite{Hsieh:2008}\cite{Guo:2012}\cite{Cossentino:2010}, as in Figure~\ref{holons2}. Such systems are, at the same time, both parts and wholes within a holarchy, thus giving a formal characterization to the notion of Holon (Figure~\ref{partVsWhole} and~\ref{partVsWhole2}). 

The paper presents an explicit approach for the recursive modeling of systems.  The approach forces the modeler to expressing the behavior of composition by a single state machine, called whole.  Such a state machine plays the double role of being both an executable specification of the behavior of the system, and to be an interface for further composition of the entire assembled system. 
This double side, ``Janus''-like feature makes such kind of systems suitable for modeling, as observed, the behavior of Holons.

The explicit approach may be used in order to partition safety tasks into hierarchically arranged modules, each checked incrementally.  Real-time critical systems, for example, may benefit from the approach since it allows to decompose a single, monolithic, control program into smaller, safe, reusable and composable systems.  It is for example possible to defeat the overall complexity issues given by the concurrent modeling of operating modes and failure management policies.  For example,
fail silently sub-devices may be used as components for assembling a device behavior, which is able, at the higher level to \emph{reduce}
the fail silent behavior to a more tractable 
fail explicit behavior.  The latter, in turn, may be used, at the next composition level, to obtain a fail safe or fail operational behavior.  An example of such hierarchical arrangement of failure modes is given in~\cite{pazziWHCM10}.

\bibliographystyle{eptcs}
\bibliography{pazziAiSoS2final}

\end{document}